\documentstyle[12pt,epsf,eufrak]{article}								 
\begin{document} 
\begin{center}
{\large\bf  Pair production of doubly heavy baryons}\\
\vspace*{4mm}
V.V.Braguta and A.E.Chalov,\\[2mm]
Moscow Institute of Physics and Technology,
Dolgoprudny, Moscow region, 147000 Russia
\end{center}
\vspace*{2mm}
\begin{abstract}
We analytically calculate the total and differential cross sections for the
pair production of doubly heavy baryons in the framework of diquark model. The
processes of  electron-positron and quark-antiquark annihilations are
considered. The fractions of doubly heavy baryons in the yields of heavy quarks
are evaluated numerically.
\end{abstract}

\section{Introduction}
High luminosities of $B$-factories and hadron colliders open a real
experimental possibility to observe doubly heavy baryons $\Xi_{QQ'}$ and
$\Omega_{QQ'}$. This prospective stimulates the theoretical interest to study
the physics of such the baryons: the spectroscopy in the framework of both
potential models \cite{pm} and QCD sum rules \cite{sr}, the lifetimes and
inclusive decay modes \cite{lt}, the mechanism of production in various
collisions and the rate of yield at accelerators \cite{prod}.

The pair production can be essential at the energies close to the threshold,
that we consider in the present paper. The physical approximation for the
calculations are caused by the apparent diquark structure of doubly heavy
baryons, since the diquark size is essentially less than the radius of
confinement determining the motion of light quark inside the $QQ'q$ system. We
suppose that, at first, the calculations of heavy diquark production should be
done, and further, the models of diquark fragmentation into the baryon
\cite{frag} can be applied.

In paper \cite{1} the differential and total cross sections for exclusive
production of meson pairs in $e^+ e^- $ annihilation was calculated in the
franework of constituent quark model. The results were obtained  for the case
of final particles in pseudoscalar and vector states close to the threshold.
Following the same procedure, we examinate the processes of $e^+e^-\rightarrow
{\EuFrak d} \bar {\EuFrak d}$ and $q \bar q \rightarrow {\EuFrak d} \bar
{\EuFrak d}$, where ${\EuFrak d}$, $\bar{\EuFrak d}$ are diquark and
antidiquark, respectively (in our calculations we neglect masses of
annihilating particles). Differential and total cross sections for the
exclusive production of heavy diquark pairs in axial-axial, axial-scalar and
scalar-scalar states are determined for the diquark composed of different heavy
quarks. We also consider the case of axial diquark composed of equivalent
quarks. We do not concern for the annihilation into the pseudoscalar and vector
diquarks, since their production does not contribute in the leading order of
$1/m$ expansion. If one supposes the fragmentation of diquarks into the doubly
heavy baryons, the formulae derived may be useful in the calculation of cross
sections for the pair production of baryons.

In section 2, we describe basic points of the constituent quark model. Section
3 and 4 contains matrix elements, differential and total cross sections for
the processes of $e^+e^-$  and $q \bar q$ annihilation into the pairs of
diquarks. In section 5 numerical results are given. In Conclusion we summarize
the consideration.

\section{Basic points of the model}

In this paper we use the constituent quark model \cite{1}, which considers the
quark masses and leptonic constants as the only input parameters.

According to the model the diquark ${\EuFrak d}=(Q_{1}Q_{2})$ mass is equal to
$$M=m_{1}+m_{2}.$$

Four-momenta of the quarks entering the diquark are given by
$$
k_{Q_1} = \frac {m_1} {M} P+q,
$$
$$
k_{Q_2} = \frac {m_2} {M} P-q,
$$
where $P$ is the meson momentum.

To represent The Fock state of diquark with two different quarks in the model,
we use the principle of superposition for the wave packages with distributions
$\Psi$. Therefore, we find the following:

in the case of scalar diquark
\begin{equation}
|S_{\EuFrak d}^i\rangle=\frac{\epsilon_{ijk}}{\sqrt{2}}\int \frac{d^3q}{(2\pi
)^3}\Psi_s (q)
\sum_{\lambda_1\lambda_2} \frac {(\Psi_{\lambda_1}^{\dagger}\hat
C\gamma_5\Psi_{\lambda_2})
^*} {\sqrt{2}} \hat a_{\lambda_1}^{j+} \hat b_{\lambda_2}^{k+} |0\rangle,
\end{equation}

and for the axial diquark
\begin{equation}
|A_{\EuFrak d}^i\rangle=\frac{\epsilon_{ijk}}{\sqrt{2}}\int \frac{d^3q}{(2\pi
)^3}\Psi_a (q)
\sum_{\lambda_1\lambda_2} \frac {(\Psi_{\lambda_1}^{\dagger}\hat C \gamma_{\mu}
\Psi_{\lambda_2})
^*} {\sqrt{2}} e_{\mu} \hat a_{\lambda_1}^{j+} \hat b_{\lambda_2}^{k+}
|0\rangle,
\end{equation}
where $\hat{C}$ is a matrix of charge conjugation, $i$ and $j$ are the colour
indices, $\hat {e}=e_{\mu} \gamma_{\mu}$, $ e_{\mu} $ is a polarisation vector
of axial diquark, $\hat a$ and $\hat b$ denote the operators of quark creation,
so that the axial and scalar diquarks in the Fock space are normilized in the
following way:
\begin{eqnarray}
\langle S^i_{\EuFrak d}(P)|S^j_{\EuFrak d}(P')\rangle &=& (2\pi)^3  \delta_{ij}
\delta (\vec P-\vec P'),
\\
\langle A^i_{\EuFrak d}(P,\lambda)|A^j_{\EuFrak d}(P',\lambda')\rangle &=&
(2\pi)^3
\delta_{ij} 
\delta_{\lambda \lambda'} \delta (\vec P-\vec P'),
\end{eqnarray}
$ \lambda $ and $ \lambda' $ are diquark polarisation indices.

If the heavy quarks are identical, then the Pauli principle must be taken into
account. So, the above formulae have to be divided by $\sqrt{2}$ and
antisymmetrized over the permutaions of $\hat a$ and $\hat b$ operators. 

The heavy quark propagator has the form 
$$
S(k)=( k_{\mu}\gamma_{\mu}+m )D(k),
$$
where
$$
D^{-1}(k)=k^2-m^2.
$$

\section{ Amplitudes and cross sections for the $e^+e^-$ annihilation}

In this section we consider the $e^+e^-$ annihilation into the diquark and
antidiquark pairs. The diagrams, which contribute into the processes in the
leading order, are shown in Figs. 1a and 1b. The colour factor for the
processes involved is equal to 
$$
{\rm Colour}_{ij}= -\frac {2} {3} \delta_{ij}~,
$$
where $i$, $j$ are the colours indices of diquark and antidiquark,
correspondingly.

\subsection{Annihilation into pairs of scalar diquarks}

The matrix element for the pair production of  scalar diquarks may be writen as
follows:
\begin{equation}
{\cal M}_{ss}=-i\frac {64 \pi^2 } {3}~ \frac {f_{ss} } {s^2}
~\delta_{ij} ~
|\Psi_{s} (0)|^2 ~ (P'  _{\mu}  -P_{\mu}) l_{\mu}
\end{equation}
where $ f_{ss} $ is equal to
\begin{eqnarray}
f_{ss} &=& M \left( \alpha_s \Bigl(\frac {m_1^2}{M^2}s \Bigr)\cdot \frac {q_2 }
{m_1^2} + 
\alpha_s\Bigl(\frac {m_2^2}{M^2}s\Bigr)\cdot
\frac {q_1 }
{m_2^2}\right)\alpha_{em} \left(s \right)-
\nonumber \\ &&
-\frac {2M^3}{s}
\left(\alpha_s\Bigl(\frac {m_1^2}{M^2}s \Bigr)\cdot\frac {q_2 m_2} {m_1^3} + 
\alpha_s \Bigl(\frac {m_2^2}{M^2}s \Bigr)\cdot \frac {q_1m_1} {m_2^3}\right)
\alpha_{em} \left(s \right),
\end{eqnarray}
and $l_{\mu}$ denotes the leptonic vector current, $ \Psi_s (0) $ is a wave
function of scalar diquark at the origin. The charges of the quarks $Q_1$ and
$Q_2$ are equal to $q_1$ and $q_2$. $P'$, $P$  are the four-momenta of scalar
diquark and antidiquark, correspondingly.

After simple algebraic calculations for the differential cross section 
$ d \sigma_{ss} $/ ${d \cos \theta} $ we get the following expression:
\begin{equation}
\frac {d \sigma_{ss}} {d \cos \theta}=64 \pi^3 \frac {f_{ss}^2 }
{3s^3}~|\Psi_s(0)|^4
\left(1-\frac {4M^2} {s}\right)^{3/2} (1-\cos^2\theta),
\end{equation}
where $\theta $ is the angle between the momenta of lepton and diquark.

The total cross section for the exclusive production of heavy scalar diquark
pairs in $e^+e^-$ annihilation is equal to
\begin{equation}
\sigma_{ss}=256 \pi^3 \frac {f_{ss}^2} {9 s^3}~|\Psi_s(0)|^4	
\left(1-\frac {4M^2} {s}\right)^{3/2}.
\end{equation}

\subsection{ Annihilation into axial and scalar diquarks}

The matrix element for the pair production of axial antidiquark and scalar
diquark may be writen as follows:
\begin{equation}
{\cal M}_{as}=-\frac {128 \pi^2} {3s^3} \delta_{ij} f_{as}
\Psi_s^* (0) \Psi_a (0) \epsilon_{\mu \alpha \beta \gamma} e_{\alpha} P_{\beta}
q_{\gamma} l_{\mu},
\end{equation}
where $ f_{as} $ is	equal to
$$
 f_{as}=M^3\left(\alpha_s\Bigl(\frac {m_1^2}{M^2}s \Bigr)\cdot \frac {q_2}
 {m_1^3}-
 \alpha_s\Bigl(\frac {m_2^2}{M^2}s \Bigr)\cdot
 \frac {q_1} {m_2^3} \right)\alpha_{em} \left(s \right).
$$
$ \Psi_a (0) $ is a wave function of axial diquark at the origin, $~q=P+P'$.

Carring out algebraic calculations for the differential cross
section  $ d \sigma_{as}/{d \cos \theta} $	we find the following
expression:
\begin{equation}
\frac {d \sigma_{as}} {d \cos \theta}=64 \pi^3~ \frac{f_{as}^2 }
{3s^4 } ~|\Psi_s(0) \Psi_a(0)|^2  \left(1-\frac {4M^2} {s}\right)^{3/2}
(2-\sin^2 \theta).
\end{equation}

The total cross section for the exclusive production of heavy scalar diquark
and axial antidiquark pairs in $e^+e^-$ annihilation is	given by the following
formula:
\begin{equation}
\sigma_{as}= 512 \pi^3~ \frac{f_{as}^2 } {9 s^4 }~
|\Psi_s(0) \Psi_a(0)|^2  \left(1-\frac {4M^2} {s}\right)^{3/2}.
\end{equation}
In the above formulae the difference between the masses of axial and scalar 
diquarks is neglected.

\subsection{ Annihilation into pairs of axial diquarks}

The matrix element for the exclusive pair production of two axial diquarks 
may be represented as follows:
\begin{equation}
{\cal M}_{aa}= -i\frac {128 \pi^2 } {3s^3} \delta_{ij}
~|\Psi_a
(0)|^2 \left(f^{[1]}_{aa} (P'_{\mu}-P_{\mu})
(e'^* e)+f^{[2]}_{aa}( (e'^* q)e_{\mu}-(eq)e'^*_{\mu}) \right) l_{\mu},
\end{equation}
where $f^{[1]}_{aa}$ and $f^{[2]}_{aa}$ are equal to 
\begin{eqnarray}
f^{[1]}_{aa} &=& M^3 \left(\alpha_s\Bigl(\frac {m_1^2}{M^2}s \Bigr)\cdot \frac
{q_2 m_2} {m_1^3}+
\alpha_s\Bigl(\frac {m_2^2}{M^2}s \Bigr)\cdot \frac{q_1 m_1} {m_2^3} \right)
\alpha_{em} \left(s \right),
\\ 
f^{[2]}_{aa} &=& M^4 \left(\alpha_s\Bigl(\frac {m_1^2}{M^2}s \Bigr)\cdot \frac
{q_2} {m_1^3}+ \alpha_s\Bigl(\frac {m_1^2}{M^2}s \Bigr)\cdot \frac {q_1}{m_2^3}
\right) \alpha_{em} \left(s \right).
\end{eqnarray}
For the differential cross section $d \sigma_{aa}/{d \cos \theta} $ we
calculate the following expression:
$$
\frac {d \sigma_{aa}} {d \cos \theta}=
\frac {512 \pi^3 } {3 s^5}~|\Psi_a(0)|^4~
\left(1-\frac {4M^2} {s}\right)^{3/2}({\cal A}-{\cal B} \cos^2 \theta),	   
$$
where  ${\cal A}$ and ${\cal B}$ have the form 
$$
{\cal A}=(f^{[1]}_{aa})^2 (8+(\eta-2)^2 )-2 f^{[1]}_{aa} f^{[2]}_{aa}
\eta(\eta-2)+
(f^{[2]}_{aa})^2(\eta^2+2\eta),
$$
$$
{\cal B}= (f^{[1]}_{aa})^2 (8+(\eta-2)^2 )-2f^{[1]}_{aa}f^{[2]}_{aa}
\eta(\eta-2)+
(f^{[2]}_{aa})^2(\eta^2-2\eta),
$$
and $\eta=s/M^2 $ .

The total cross section for the exclusive production of heavy axial diquarks
pairs in $e^+e^-$ annihilation is given by
\begin{equation}
\sigma_{aa}= \frac {1024 \pi^3 } {9 s^5} ~|\Psi_a(0)|^4~ 
\left(1-\frac {4M^2} {s}\right)^{3/2}(3{\cal A}-{\cal B} ).
\end{equation}
Actually, the behaviour of this cross section at large $s$ is $\sigma_{aa}\sim
1/s^3$.

\subsection { The diquark containing two equivalent particles }

If the diquarks are composed of identical particles, the above formulae must be
changed. Evidently, we have to take into account the production of axial
diquark, only. Since the smallest angular momentum for the scalar diquark is
equal to unit, it does not contribute in the leading order of $1/m$ expansion.
In contrast, the angular momentum of axial diquark can be equal to zero.

All formulae writen down for the annihilation into two axial diquarks remains
valid except two form factors $f^{[1]}_{aa},f^{[2]}_{aa}$. They must be
transformed in
the following way:
\begin{eqnarray}
f^{[1]}_{aa} &=& 2q \alpha_s(M^2)M\; \alpha_{em} \left(s \right), 
\\
f^{[2]}_{aa} &=& 4q\alpha_s(M^2)M\; \alpha_{em} \left(s \right),
\end{eqnarray}
where q is the quark charge.

\section{Amplitudes and cross sections for the $q \bar q$ annihilation}

In this section we consider the $q \bar q$ annihilation into the diquark and
antidiquark pairs. The diagrams, which contribute in the leading order, are
shown in  Figs. 1a, 1b, 2, 3a and 3b. The diagram shown in Fig. 2 results in
zero in our approach,  when the quarks are on the mass shells.

The colour factor for the diagrams shown in  Figs. 1a, 1b is
equal to
$$
{\rm Colour}^{[1]}_{(ij)(lm)}=\frac {1}{3} t^a_{ij} t^a_{lm},
$$
where $m,l$ are the colour indices of annihilating quark and antiquark,
correspondingly. It is easy to notice that this is a colour octet state. The
colour factor for the diagrams shown in Figs. 3a, 3b is equal to
$$
{\rm Colour}^{[2]}_{(ij)(lm)}=\frac {5}{12} t^a_{ij} t^a_{lm}-\frac
{1}{9}\delta_{ij}\delta_{lm},
$$
that appears to be a mixture  of octet and singlet colour states.

\subsection{Annihilation into pairs of scalar diquarks}

The matrix element for the exclusive pair production of two scalar diquarks 
may be written as follows:
\begin{equation}
{\cal M}_{ss}=\frac {32 \pi^2 i } {s^2} ~|\Psi_s(0)|^2
\left(\tilde f^{[1]}_{ss}\frac {2t^a_{if}t^a_{lm}}{3}P'_{\mu}-
\tilde f^{[2]}_{ss}\left(\frac
{5t^a_{if}t^a_{lm}}{6}-\frac{2\delta_{if}\delta_{lm}}{9}\right)
\frac {\left(p,P'-P\right)} {s}P_{\mu}\right)l_{\mu},
\end{equation}
where  $\tilde f^{[1]}_{ss}$ and $\tilde f^{[2]}_{ss}$ are equal to
\begin{eqnarray}
\tilde f^{[1]}_{ss} &=& M \left(\frac {\alpha_s \Bigl(\frac {m_1^2}{M^2}s
\Bigr) }
{m_1^2}
+ 
\frac {\alpha_s\Bigl(\frac {m_2^2}{M^2}s\Bigr)} {m_2^2}\right)
\alpha_{s} \left(s \right)
-\frac {2M^3}{s}
\left(\alpha_s\Bigl(\frac {m_1^2}{M^2}s \Bigr)\cdot\frac {m_2} {m_1^3} + 
\alpha_s \Bigl(\frac {m_2^2}{M^2}s \Bigr)\cdot \frac {m_1} {m_2^3}\right)
\alpha_{s} \left(s \right),
\\
\tilde f^{[2]}_{ss} &=& \frac{M^5}{m_1^3m_2^3}
\alpha_s \Bigl(\frac {m_1^2} {M^2}s \Bigr)  \alpha_s \Bigl(\frac {m_2^2} {M^2}s
\Bigr) ,
\end{eqnarray}
and $p$ is the four-momentum of annihilating quark.

For the differential cross section $ d \sigma_{ss}/{d \cos \theta} $ we get
the following expression:
\begin{eqnarray}
\frac {d \sigma_{ss}} {d \cos \theta} &=&
\frac{8\pi^3}{81s^3}~|\Psi_s(0)|^4\left(\left(2\tilde f^{[1]}_{ss}+
\frac{5}{4}\tilde f^{[2]}_{ss}\sqrt{1-\frac {4M^2}
{s}}\cos\theta\right)^2+\frac{(\tilde f^{[2]}_{ss})^2}{2}(1-\frac
{4M^2}{s})
\cos^2\theta\right)
\cdot \nonumber \\ &&
\cdot \left(1-\frac{4M^2}{s}\right)^{3/2}(1-\cos^2\theta).
\end{eqnarray}
The total cross section for the exclusive production of two heavy scalar
diquarks in the $q \bar q$ annihilation is	given by the formula
\begin{equation}
\sigma_{ss}=\frac{ 8 \pi^3}{81s^3}~|\Psi_s(0)|^4
\left(1-\frac{4M^2}{s}\right)^{3/2}\left(\frac{16}{3}(\tilde
f^{[1]}_{ss})^2+\frac{11}
{20}
\left(1-\frac{4M^2}{s}\right)(\tilde f^{[2]}_{ss})^2\right).
\end{equation}

\subsection{Annihilation into axial and scalar diquarks}

The matrix element for the pair production of  axial antidiquark and scalar
diquark may be represented in the following way:
\begin{equation}
{\cal M}_{as}=\frac {32 \pi^2}{s^3}
\Psi_s^*(0) \Psi_a(0)\left(-\tilde f^{[1]}_{as}\frac
{2t^a_{if}t^a_{lm}}{3}~\epsilon_{\mu\alpha\beta\gamma}
P_{\beta}e_{\alpha}q_{\gamma}-\tilde f^{[2]}_{as}\left(\frac
{5t^a_{if}t^a_{lm}}{6}-\frac{2\delta_{if}\delta_{lm}}{9}\right)
\epsilon_{\mu\nu\alpha\beta}q_{\alpha}e_{\beta}p_{\nu}
\right)l_{\mu},
\end{equation}
where  $\tilde f^{[1]}_{as}$and $\tilde f^{[2]}_{as}$ are equal to
\begin{eqnarray}
\tilde f^{[1]}_{as} &=& M^3 \left( \frac{ \alpha_s \Bigl(\frac {m_1^2}{M^2}s
\Bigr)}{m_1^3}-
\frac{ \alpha_s \Bigl(\frac {m_2^2}{M^2}s \Bigr) }{m_2^3} \right)\alpha_s(s),
\\
\tilde f^{[2]}_{as} &=& \alpha_s \Bigl(\frac {m_1^2}{M^2}s \Bigr) 
\alpha_s \Bigl(\frac {m_2^2}{M^2}s \Bigr) 
\frac{M^5(m_2-m_1)}{m_1^3m_2^3}.
\end{eqnarray}
The differential cross section $ d \sigma_{as} /{d \cos \theta} $ are given by
\begin{eqnarray}
\frac {d \sigma_{as}} {d \cos \theta} &=& \frac
{64{\pi}^3}{81s^4}~ |\Psi_s(0)\Psi_a(0)|^2 \sqrt{1-\frac{4M^2}{s}}\cdot
\left(\frac{(\tilde f^{[1]}_{as})^2}{2} \left(1-\frac{4M^2}{s}\right)
(1+\cos^2\theta) \right. \\ && \left. + \frac{33(\tilde
f^{[2]}_{as})^2}{16}\left(
1+\frac{s-4M^2}{8M^2}\sin^2\theta\right) 
+\frac {5}{2}\tilde f^{[1]}_{as}\tilde f^{[2]}_{as} \sqrt {1-\frac
{4M^2}{s}}\cos\theta
\right). \nonumber
\end{eqnarray}
The total cross section for the exclusive production of  heavy axial
antidiquark and scalar diquark pairs in the $q \bar q$ annihilation equals
\begin{equation}
\sigma_{as}=\frac {64{\pi}^3}{243s^4}~
|\Psi_s(0)\Psi_a(0)|^2 \sqrt{1-\frac{4M^2}{s}}
\left(4(\tilde f^{[1]}_{as})^2 \left(1-\frac{4M^2}{s}\right)
+\frac{33(\tilde f^{[2]}_{as})^2}{8}\left(2+\frac{s}{4M^2} \right)
\right).
\end{equation}
Here we have also neglected the difference between masses of axial and scalar
diquarks.

\subsection{Annihilation into pairs of axial diquarks}

The amplitude for the exclusive pair production of two axial diquarks in 
the process may be represented as follows:
\begin{eqnarray}
{\cal M }_{aa} &=& \frac {32 \pi^2 i} {3}  \frac {\alpha_s(4m_1^2)
\alpha_s(4m_2^2)} {m_1^3 m_2^3
s^3}~M^5~|\Psi_a(0)|^2\biggl\{\biggl(~(-Pe'^*)\cdot
\Bigl((pP')(le)+(lP')(pe)
\Bigr)
\nonumber \\ && +
(-P'e)\cdot\Bigl((Pp)(le'^*)+
(lP)(pe'^*)\Bigr)+(ee'^*)\cdot\Bigl((lP)(pP')+(lP')(pP)\Bigr)
\nonumber \\ && +
 \frac
 {s}{2}\cdot\Bigl((pe)(le'^*)+(le)(pe'^*)\Bigr)~\biggr)
\cdot \Bigl(t_{if}^a
t_{lm}^a-\frac {4} {15} \delta_{if}\delta_{lm} \Bigr)
\\ && +
\biggl(\tilde f_{aa}^{[2]} \Bigl((el)(Pe'^*)-(P'e)(e'^*l)
\Bigr)+\tilde f_{aa}^{[1]}(Pl)(ee'^*)
\biggr)
\cdot
\Bigl(t_{if}^a t_{lm}^a\Bigr)\biggr\}. \nonumber
\end{eqnarray}
The differential cross section $d \sigma_{aa}/{d \cos \theta} $	is given by the
following expression:
\begin{eqnarray}
  \frac {d \sigma_{aa} } {d \cos \theta} &=& \frac {200 \pi^3} {81}
  \alpha_s^2(4m_1^2) \alpha_s^2(4m_2^2)
  \frac {M^{10}} {m_1^6 m_2^6 s^7}|\Psi_a(0)|^4 \sqrt{1-\frac{4M^2} {s}}
  \nonumber \\ && \cdot
  (a_4 \cos^4{\theta}+a_3
  \cos^3{\theta}+a_2 \cos^2{\theta}+a_1 \cos{\theta}+a_0).
\end{eqnarray}
The total cross section for the exclusive production of  heavy axial diquarks
pairs in the $q \bar q$ annihilation has the form
\begin{equation}
   \sigma_{aa}=\frac {200 \pi^3} {81} \alpha_s^2(4m_1^2) \alpha_s^2(4m_2^2)
   \frac
  {M^{10}} {m_1^6 m_2^6 s^7}|\Psi_a(0)|^4  \sqrt{1-\frac{4M^2} {s}} \left(
  \frac {2} {5} a_4 
  + \frac {2} {3} a_2 +2 a_0 \right),
\end{equation}
where we have used the following definitions:
\begin{eqnarray}
\tilde f_{aa}^{[1]} &=& -\frac {\alpha_s \left( s \right)}{\alpha_s(4m_1^2)
\alpha_s(4m_2^2)} \frac {8 m_1^3 m_2^3} {5 M^2} \left ( \frac
{\alpha_s(4m_1^2)m_2}
{m_1^3}+\frac {\alpha_s(4m_2^2)m_1} {m_2^3} \right ), \\
\tilde f_{aa}^{[2]} &=&  \frac {\alpha_s \left( s \right)} {\alpha_s(4m_1^2)
\alpha_s(4m_2^2)}\frac {4 m_1^3 m_2^3} {5 M} \left ( \frac { \alpha_s(4m_1^2) }
{m_1^3}+\frac {\alpha_s(4m_2^2)} {m_2^3} \right ), \\
a_4 &=& \frac {99} {400} s^2 (s-4 M^2)^2, \\
a_3 &=& \frac {1} {8} \frac {s^3} {M^2} \left(2 \tilde f_{aa}^{[2]} s+\tilde
f_{aa}^{[1]} (6 M^2+s) \right) \left(1-4 \frac {M^2} {s}\right )^{3/2}, \\
a_2 &=& \frac {1} {16 M^4} s (s-4M^2)\biggl\{4\tilde f_{aa}^{[1]} \tilde
f_{aa}^{[2]} s (s-2M^2)+(\tilde f_{aa}^{[1]})^2(12 M^4-4 M^2 s+s^2) \nonumber
\\ && +
s\Bigl(~\frac {33} {25}(12 M^6-M^4 s)+4(\tilde f_{aa}^{[2]})^2
(s-2M^2)~\Bigr)\biggr\}, \\
a_1 &=& -\frac {s^2} {8 M^2} \sqrt {1-4 \frac {M^2} {s}} \Bigl(~2 \tilde
f_{aa}^{[2]} s (s+4 M^2)+\tilde f_{aa}^{[1]}(-24M^4+2M^2s+s^2)~\Bigr), \\
a_0 &=& -\frac {s} {16 M^4} \biggl(4\tilde f_{aa}^{[1]} \tilde f_{aa}^{[2]} s
\Bigl(8M^4-6M^2s+s^2\Bigr)+(\tilde
f_{aa}^{[1]})^2\Bigl(-48M^6+28M^4s-8M^2s^2+s^3\Bigr) \nonumber \\ &&
+2s\Bigl(~\frac{33} {25} M^4s(4M^2+s)+\tilde
f_{aa}^{[2]}(-16M^4-4M^2s+2s^2)~\Bigr)\biggr). 
\end{eqnarray}

\subsection{The diquark with two identical particles}
In this section we consider the $q \bar q$ annigilation into pair of diquarks
composed of two identical heavy quarks. So, as in $e^+e^-$ annihilation, we
have to take into account the axial diquark production, only.

The amplitude has the form
\begin{eqnarray}
{\cal M }_{aa} &=& \frac {512 \pi^2 i} {3  }  \frac {\alpha^2_s(M^2)} {M
s^3}~|\Psi_a(0)|^2\biggl\{\biggl(~(-Pe'^*)\cdot
\Bigl((pP')(le)+(lP')(pe) \Bigr) \nonumber \\ && +
(-P'e)\cdot\Bigl((Pp)(le'^*)+
(lP)(pe'^*)\Bigr)+(ee'^*)\cdot\Bigl((lP)(pP')+(lP')(pP)\Bigr) \nonumber \\ && +
 \frac
 {s}{2}\cdot\Bigl((pe)(le'^*)+(le)(pe'^*)\Bigr)~\biggr)
\cdot \Bigl(t_{if}^a
t_{lm}^a-\frac {4} {15} \delta_{if}\delta_{lm} \Bigr) \nonumber \\ && +
\biggl(\tilde f_{aa}^{[2]} \Bigl((el)(Pe'^*)-(P'e)(e'^*l)
\Bigr)+\tilde f_{aa}^{[1]}(Pl)(ee'^*)
\biggr)
\cdot
\Bigl(t_{if}^a t_{lm}^a\Bigr)\biggr\}.
\end{eqnarray}
The differential cross section $ d \sigma_{aa}/ {d \cos \theta} $ equals
\begin{equation}
  \frac {d \sigma_{aa} } {d \cos \theta}=2^{11} \frac { 25 \pi^3} {81}
  \frac {\alpha_s^4(M^2)} {M^2 s^7} |\Psi_a(0)|^4\sqrt{1-\frac{4M^2} {s}}
  (a_4 \cos^4{\theta}+a_3
  \cos^3{\theta}+a_2 \cos^2{\theta}+a_1 \cos{\theta}+a_0).
\end{equation}
The total cross section for the exclusive production is given by  
\begin{equation}
   \sigma_{aa}=2^{11} \frac {25 \pi^3} {81} 
   \frac
  {\alpha_s^4(M^2)} {M^2 s^7}|\Psi_a(0)|^4  \sqrt{1-\frac{4M^2} {s}} \left(
  \frac {2} {5} a_4 
  + \frac {2} {3} a_2 +2 a_0 \right),
\end{equation}
where we have used the following definitions:
\begin{eqnarray}
\tilde f_{aa}^{[1]} &=& -\frac {\alpha_s \left( s \right) }{\alpha_s(M^2)
} \frac { M^2 } {5 }, \\
\tilde f_{aa}^{[2]} &=& \frac {\alpha_s \left( s \right) }{\alpha_s(M^2)
} \frac { M^2 } {5 }.
\end{eqnarray}

\section{Numerical estimates}

The cross section ratios of the exclusive heavy diquark pair production to the
respective heavy quark cross section in $e^+ e^-$  annihilation,
$$
\sigma(e^+ e^- \rightarrow Q \bar Q)=\frac{4 \pi \alpha^2_{em}q^2_Q }{s}
\sqrt{1-\frac{4 m^2_Q}{s}} \left( 1+\frac{2 m^2_Q}{s} \right),
$$
are shown in Figs. 4a, 4b.

We see that the most optimistic expectations for the pair production is given
for the $cc$-diquarks at $B$-factories, when the hogh luminosities can yield
several thousands pairs of doubly charmed baryons.

The cross section ratios of the exclusive heavy diquark pair production to the
respective heavy quark cross section in $q \bar q$  annihilation,
$$
\sigma(q \bar q \rightarrow Q \bar Q)=\frac{8 \pi \alpha^2_{s} }{27s}
\sqrt{1-\frac{4 m^2_Q}{s}} \left( 1+\frac{2 m^2_Q}{s} \right)
$$
are shown in Figs. 4c, 4d.

Figs. 4a, 4c represent the case of diquark composed of different quarks
and Figs. 4b, 4d give the case of diquark composed of identical quarks. We put
\begin{eqnarray}
m_c &=& 1.55 ~GeV, \nonumber \\
m_b &=& 4.9 ~GeV, \\
\Lambda_{QCD} &=& 0.2 ~GeV. \nonumber
\end{eqnarray}
The values of diquark wave functions at the origin have been taken from
\cite{pm}.

In the quark-antiquark annihilation the pair production of doubly charmed
baryons gives about $10^{-5}$ fraction of charm yield at $\sqrt{s} < 100$ GeV
in the hadron collisions, while at higher energies the gluon-gluon fusion will
dominate. In experoments at fixed target the threshold behaviour results in an
additional suppression.

The inclusive production of doubly heavy baryons was considered in \cite{prod},
so that we see that the pair production results in the suppression factor about
0.1.
 
\section{Conclusion}
In this paper we have considered the exclusive production of doubly heavy
diquark pairs for the axial-axial, scalar-scalar and axial-scalar states in the
framework of the constituent quark model. The matrix elements, differential and
total cross sections for the processes of $e^+e^-$ and $q \bar q$ annihilation
have been given. We have calculated the pair production of diquarks composed of
identical heavy quarks. The obtained formulae can be used in the calculation of
cross sections for the diquark fragmentation into the doubly heavy baryons in
the processes of $e^+e^-$ annihilation and proton-antiproton inelastic
collisions.

We have found that the yield of doubly heavy baryon pairs could really reach
$10^3\div 10^6$ depending on the energies and luminosities of accelerators,
$e^+e^-$ colliders and fixed target experiments with hadron beams.

The authors thank Dr. V.V.Kiselev, who asked us for the calculations of pair
production of doubly heavy diquarks and explained some points in the problem.

This work is in part supported by the Russian Foundation for Basic Research,
grants 99-02-16558 and 00-15-96645.

\newpage
\section*{Figure Captions}

\begin{itemize}
\item[Fig. 1.]
{The diagrams of $e^+e^-$ ($q\bar q$) annihilation into the pair of doubly
heavy diquarks due to the single photon (gluon) exchange.}
\item[Fig. 2.]
{The $q\bar q$ annihilation into the diquarks due to the three-gluon
interaction.}\item[Fig. 3.]
{The diagrams of double gluon annihilation of light quarks into the
pair of doubly heavy diquarks.}
\item[Fig. 4a.]
The ratios of total cross sections for the exclusive production of $bc$ diquark
pair and the $c \bar c$ production in $e^+ e^-$ annihilation for the
axial-axial, axial-scalar and scalar scalar states.
\item[Fig. 4b.]
The ratios of cross sections for the exclusive production of $bb$ and $cc$
diquark pair and the respective $b \bar b$ and $c \bar c$ production in
$e^+ e^-$ annihilation.
\item[Fig. 4c.]
The ratios of cross sections for the exclusive production of $bc$ diquark pair
and the $c \bar c$ production in $q \bar q$ annihilation. The notations are the
same as in Fig. 4a.
\item[Fig. 4d.]
The ratios of cross sections for the the exclusive production of $bb$ and $cc$
diquark pair and the respective $b \bar b$ and $c \bar c$ production in
$q \bar q$ annihilation.
\end{itemize}

\newpage
\setlength{\unitlength}{1mm}
\begin{figure}[ph]
\begin{picture}(150, 200)
\put(-10,160){\epsfxsize=8cm \epsfbox{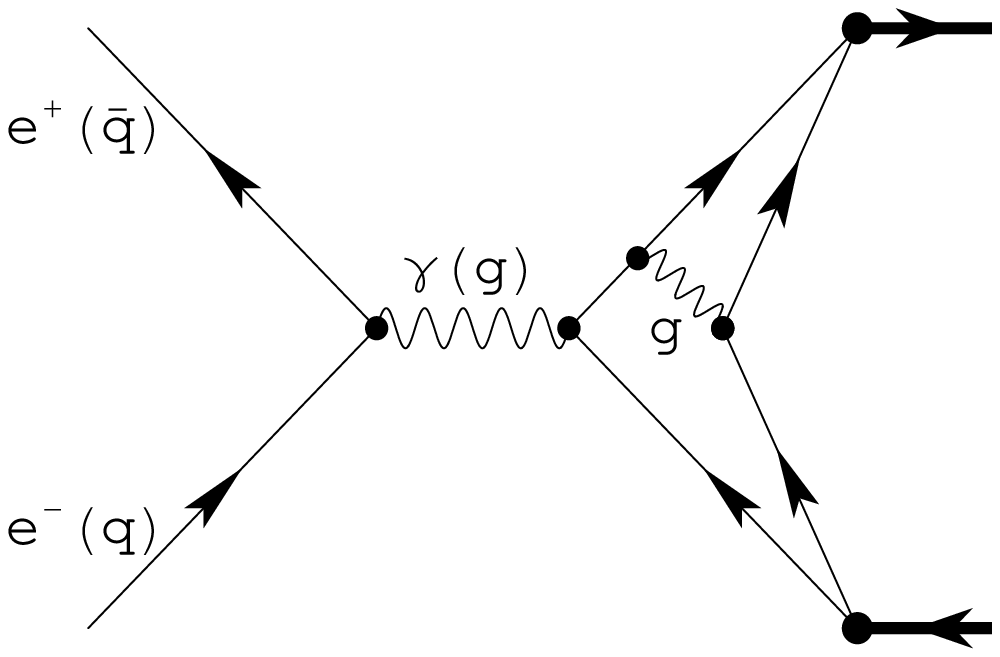}}
\bf
\put(23,160){Fig. 1a} 
\put(80,160){\epsfxsize=8cm \epsfbox{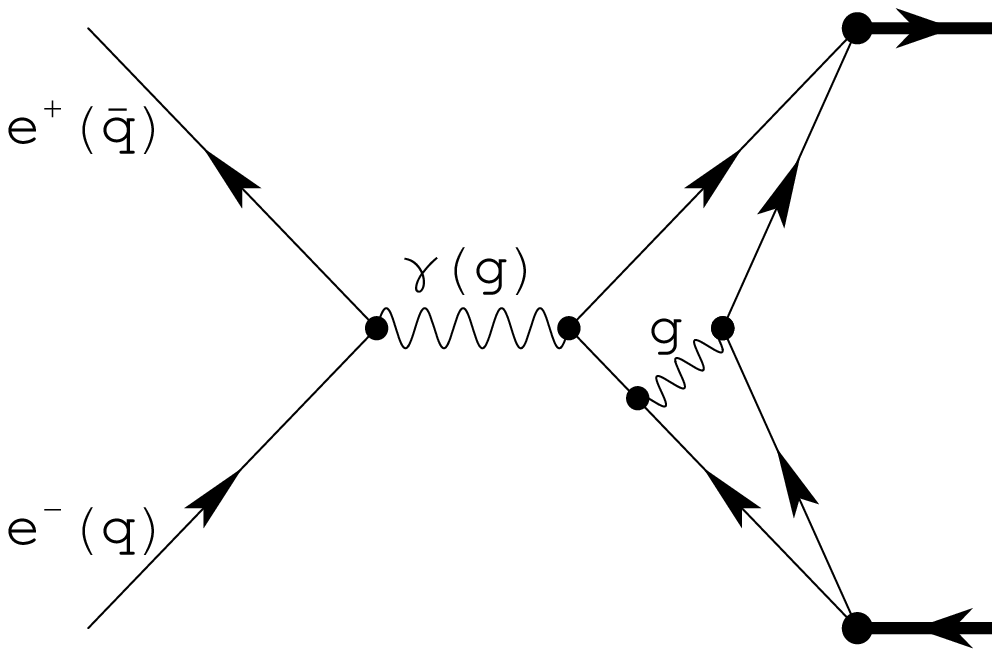}}
\put(113,160){Fig. 1b}
\put(35,90){\epsfxsize=8cm \epsfbox{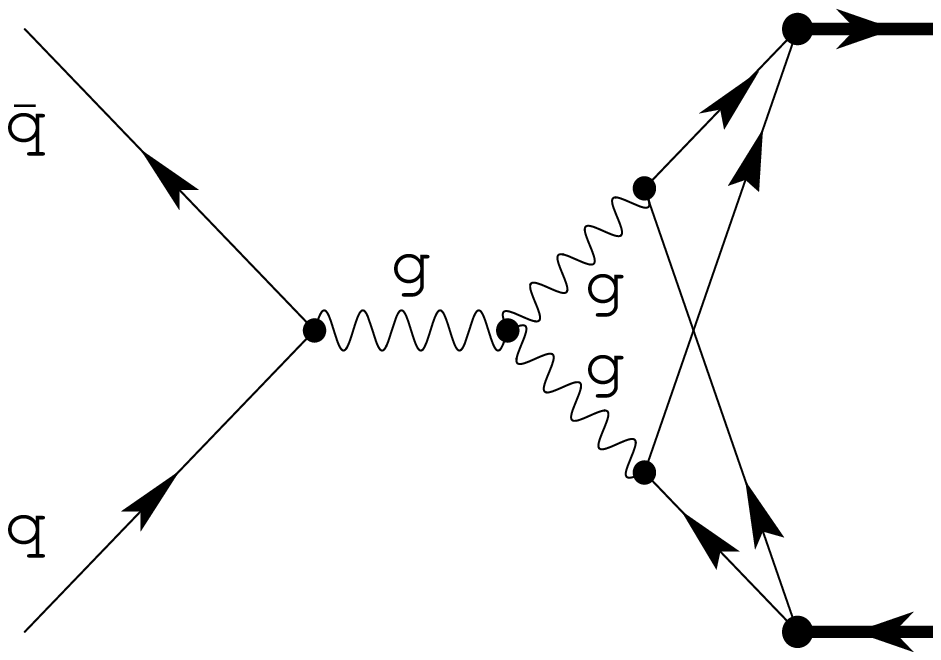}}
\put(68,90){Fig.2}
\put(-20,20){\epsfxsize=9cm \epsfbox{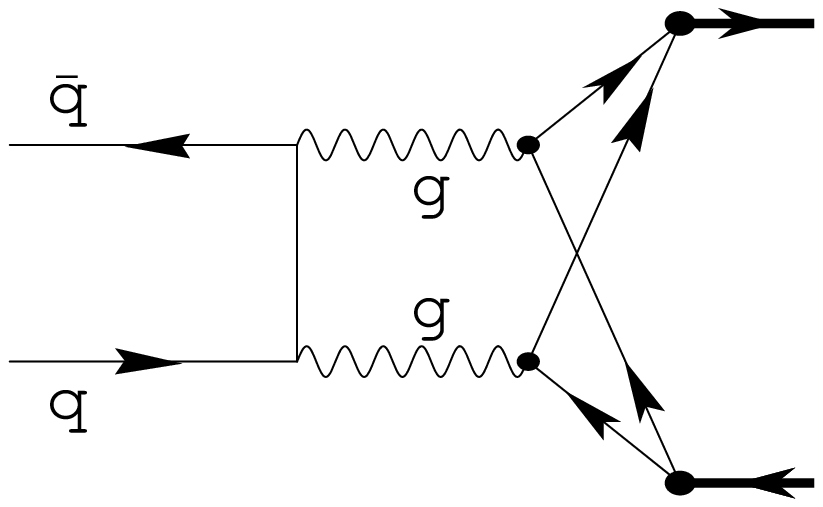}}
\put(23,20){Fig. 3a}
\put(70,20){\epsfxsize=9cm \epsfbox{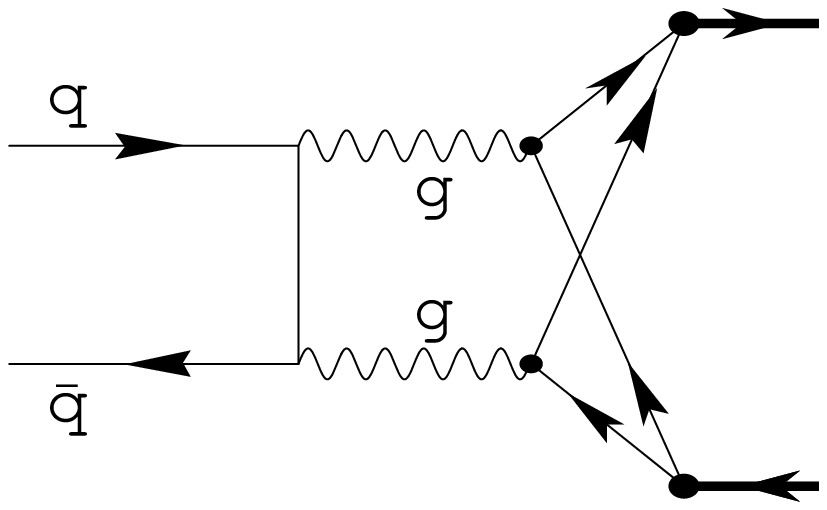}}
\put(113,20){Fig. 3b}
\put(65,210){$\EuFrak d$}
\put(65,172){$\bar{\EuFrak d}$}
\put(155,210){$\EuFrak d$}
\put(155,172){$\bar{\EuFrak d}$}
\put(110,141){$\EuFrak d$}
\put(110,103){$\bar{\EuFrak d}$}
\put(65,60){$\EuFrak d$}
\put(65,32){$\bar{\EuFrak d}$}
\put(155,60){$\EuFrak d$}
\put(155,32){$\bar{\EuFrak d}$}
\end{picture}
\end{figure}

\newpage
\setlength{\unitlength}{1mm}
\begin{figure}[ph]
\begin{picture}(150, 200)

\put(0,160){\epsfxsize=11cm \epsfbox{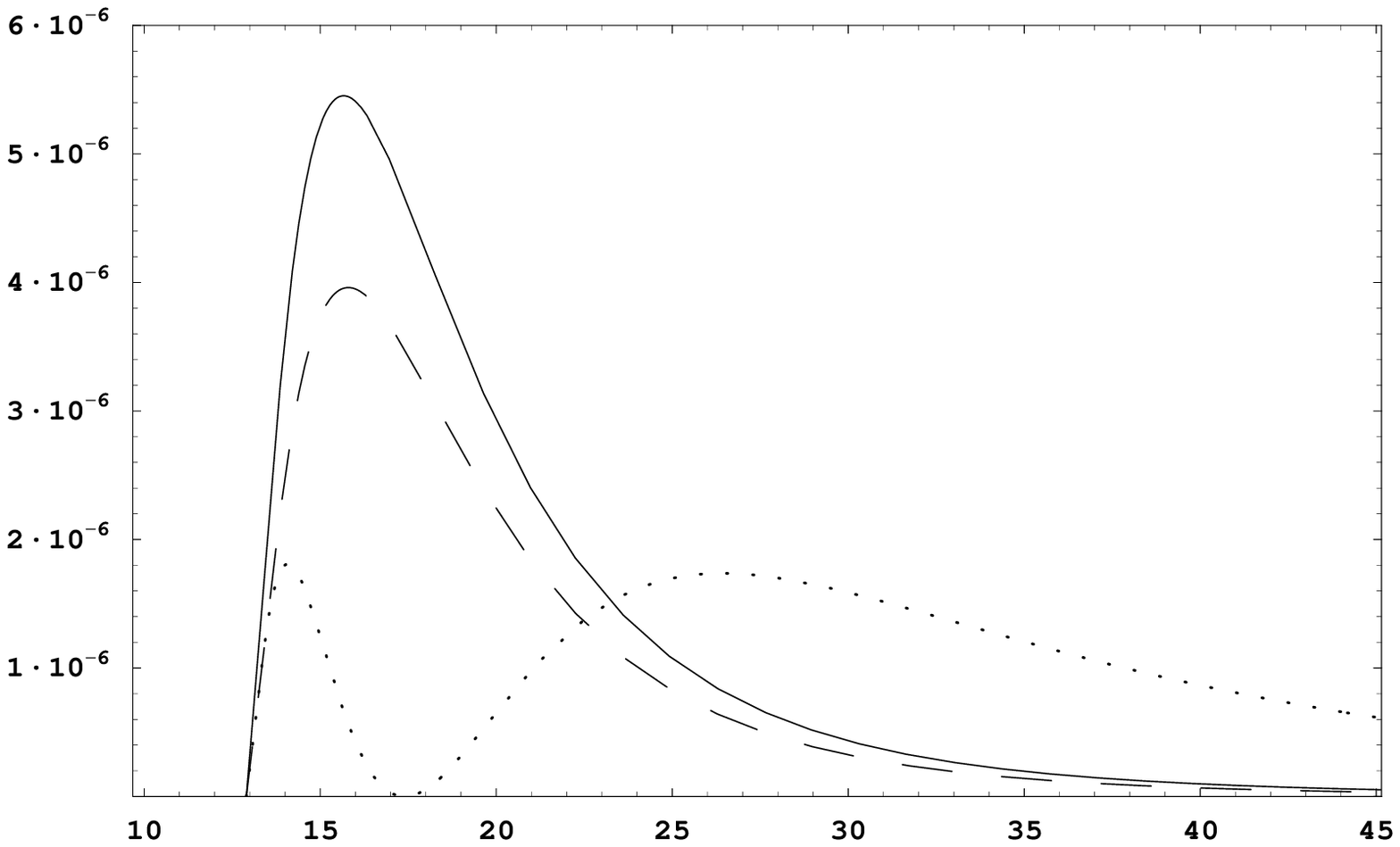}}
\put(110,165){$\sqrt{s},GeV$}
\put(10,230){$R$}
\put(70,215){-- -- -- $ 15 \cdot R_{Ax-Sc}$}
\put(70,220){------- $ R_{Ax-Ax}$ }
\put(70,211){........}
\put(80,210){ $400 \cdot R_{Sc-Sc}$}
\put(-3,155){ \bf
Fig. 4a. }

\put(-2,40){\epsfxsize=11.1cm \epsfbox{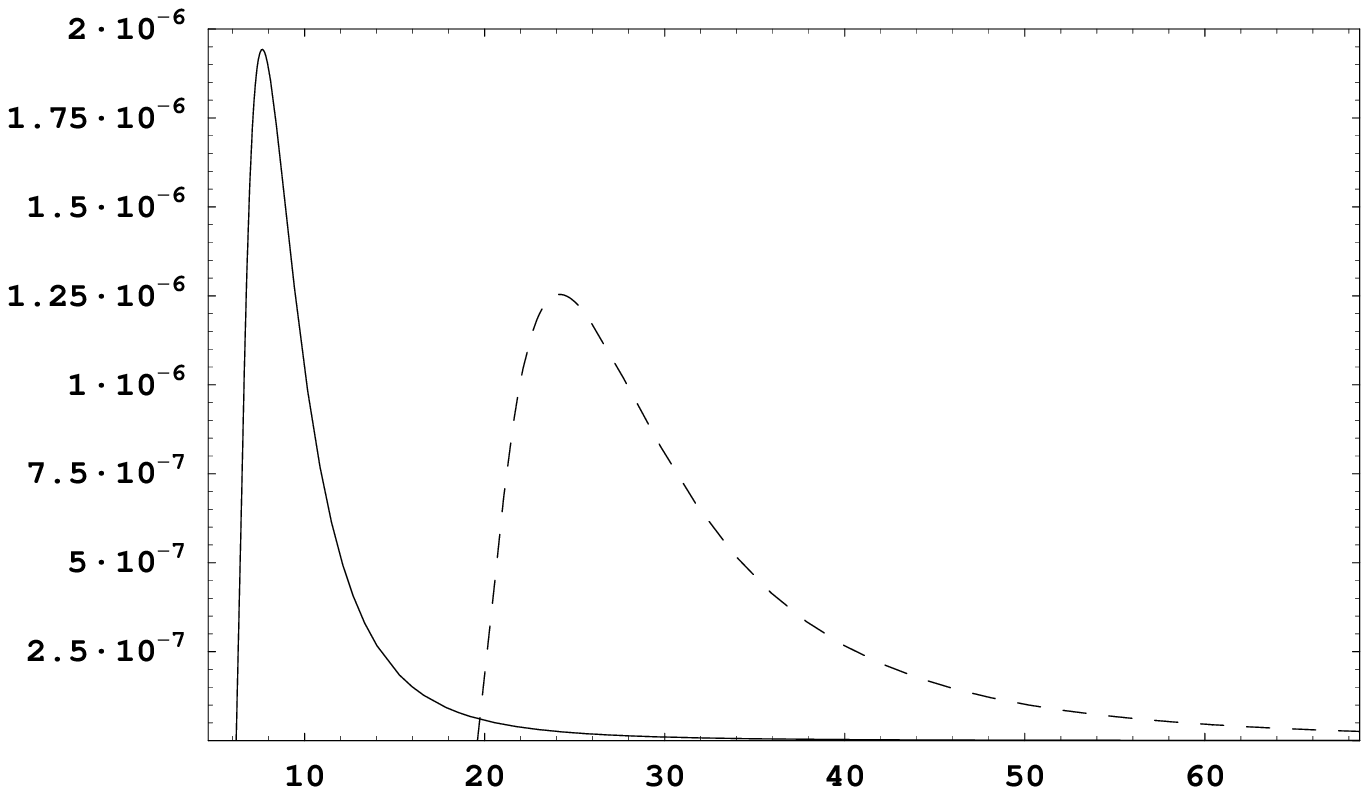}}
\put(110,45){$\sqrt{s},GeV$}
\put(10,110){$R$}
\put(80,90){-- -- -- ~~$30 \cdot R_{bb}$}
\put(80,95){------- ~~$R_{cc}$ }
\put(-3,35){ \bf Fig. 4b. }

\end{picture}
\end{figure}

\newpage
\setlength{\unitlength}{1mm}
\begin{figure}[ph]
\begin{picture}(150, 200)

\put(0,160){\epsfxsize=11cm \epsfbox{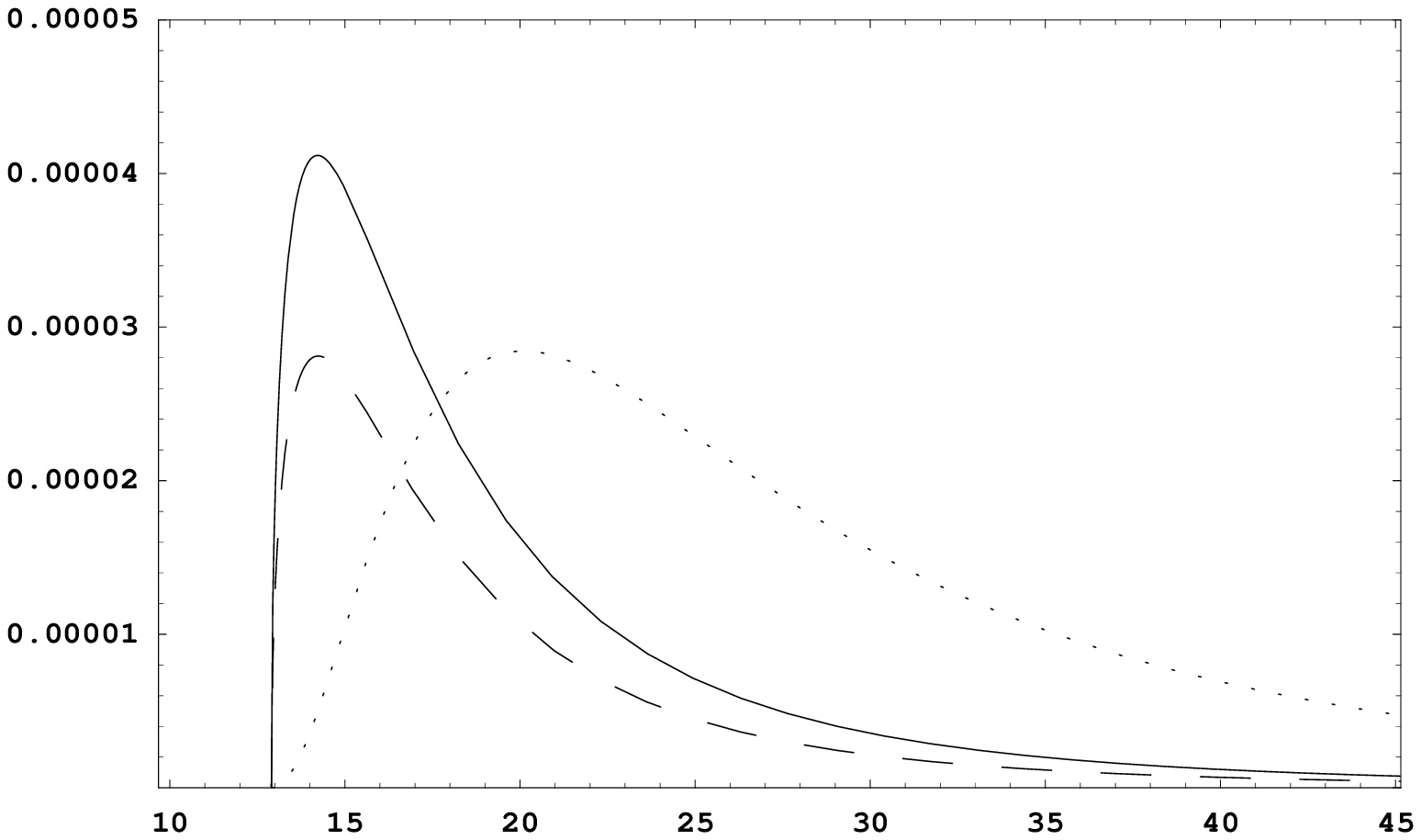}}
\put(110,165){$\sqrt{s},GeV$}
\put(10,230){$R$}
\put(70,215){-- -- -- $5 \cdot R_{Ax-Sc}$}
\put(70,220){------- $R_{Ax-Ax}$ }
\put(70,211){........}
\put(80,210){$100 \cdot R_{Sc-Sc}$}
\put(-3,155){ \bf Fig. 4c. }

\put(-2,40){\epsfxsize=11cm \epsfbox{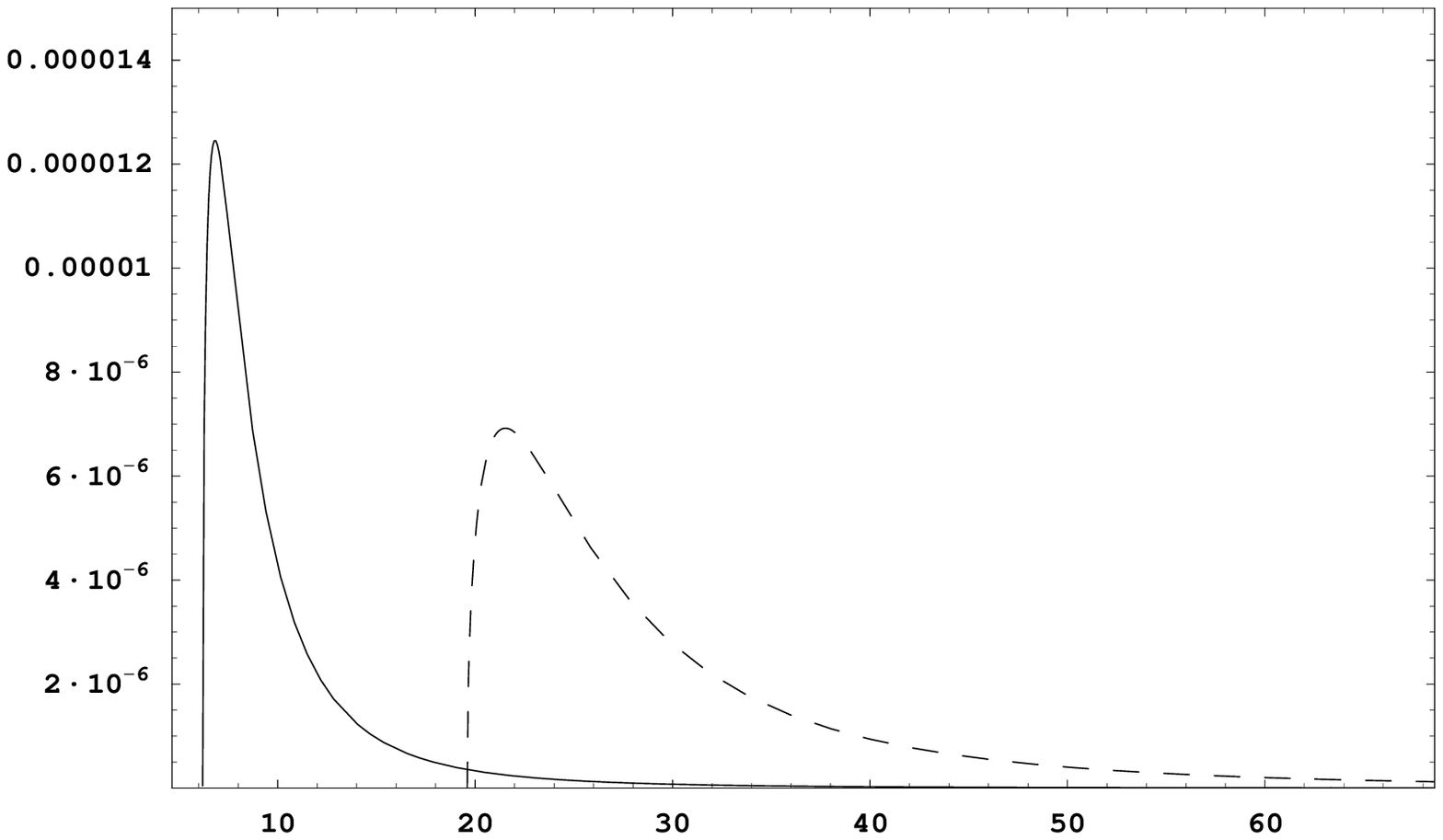}}
\put(110,45){$\sqrt{s},GeV$}
\put(10,110){$R$}
\put(80,90){-- -- --    ~~$30 \cdot R_{bb} $}
\put(80,95){-------    ~~$R_{cc}$ }
\put(-3,35){ \bf Fig. 4d. }

\end{picture}
\end{figure}

\end{document}